\documentclass[12pt]{article}
\newcommand{\be}{\begin{equation}}
\newcommand{\ee}{\end{equation}}
\newcommand{\bear}{\be\begin{array}}
\newcommand{\bea}{\begin{eqnarray}}
\newcommand{\eea}{\end{eqnarray}}

\date{May 14, 2002}
\author{G.L.~Kotkin$^{1)}$, V.G.~Serbo$^{1)}$, and V.I.~Telnov$^{2)}$\\
{\it $^{1)}$Novosibirsk State University, 630090, Novosibirsk, Russia}\\
{\it $^{2)}$Budker Institute of Nuclear Physics, 630090,
Novosibirsk, Russia}}
\title{Electron (positron) beam polarization by Compton scattering on 
 circularly  polarized laser photons~\footnote{Talk at 9-th Intern. 
Workshop on Linear Colliders (LC02), Feb. 4-8, 2002, SLAC, Stanford,
USA. To be published in Phys. Rev. ST Accelerators and Beams.}  }

\begin{document}

\maketitle

\begin{abstract}
In a number of papers an attractive method of laser polarization
of electrons (positrons) at storage rings or linear colliders has
been proposed. We show that these suggestions are incorrect and
based on errors in the simulation of multiple Compton scattering and
in the calculation of the Compton spin-flip cross sections. We argue
that the equilibrium polarization in this method is zero.
\end{abstract}

\section{Introduction}

Experiments at SLC have shown great potential of polarized
$e^{\pm}$ beams for investigation of new physical phenomena. In
all projects of future $e^+e^-,\; e^-e^-$, $\gamma \gamma$ and
$\gamma e$ linear colliders~\cite{Col},  electron and positron
beams with high degree of polarization are foreseen, though this
is not an easy task. That is why any new methods for obtaining
polarized $e^{\pm}$ beams are very welcome.

There are two well--known and recognized methods for production of
polarized beams for linear colliders. In the first method,
electron beams with a polarization of 80~\% (maybe even higher)
are obtained using photoguns~\cite{phguns}. Another method of
polarization, suitable both  for electron and positron beams, is
based on a two-step scheme~\cite{balakin}. At the first step, the
unpolarized electron beam passes through a helical undulator (or
collides with circularly polarized laser light) and produces
photons with  maximum energy of about $30\div 50$ MeV. These
photons have a high degree of circular polarization in the high
energy part of the spectrum. Then these photons pass through a thin tungsten
target and produce $e^+e^-$ pairs. At the maximum energies, these
particles have a high degree of longitudinal polarization. The
expected polarization of electron and positron beams in this
method is $45\div 60$~\% \cite{TESLA,Omori}.

Additionaly, in a number of papers a new attractive method for
production of polarized electron\footnote{This method is
applicable both for electrons and positrons, for brevity we will
omit ``positrons'' hereafter} beams, based on the process of
multiple Compton scattering of ultra-relativistic electrons on the
circularly polarized laser photons, was discussed. Here one should
distinguish between two possibilities: polarization of the beam at the cost
of a loss in intensity and polarization without loss of intensity.
The latter case is the subject of the present paper.

It is well known that in  Compton scattering the electrons of
different helicities are knocked out of the beam
differently. As a consequence, after  removal of the scattered
electrons, the electron beam can get a considerable polarization
at the expense of a considerable loss of its intensity. A detailed
consideration of this method was given in~\cite{DKS79}. Though the
basic idea of this method is correct, it was not used in practice
because the losses in intensity during the polarization
process are too large.

In this paper we critically consider another proposal of beam
polarization which is based on  {\it multiple Compton
scattering of electrons on laser photons without loss of
intensity.} It implies that during a single Compton scattering the
energy loss of an electron is small and the electron scattering
angle is small as well; therefore, the scattered electrons remain
in the beam. It means that the electron energy is of the order of
$1$ GeV for laser light with a photon energy of about $1$ eV. Such
proposals were given in papers~\cite{Bes,Ros,P1,P2}. They were
cited in a number of papers (see Refs.~\cite{Clen}, for example)
and attracted attention at the Snowmass 2001 conference. It was
suggested to implement this method either on a storage ring (where
the electron beam  collides with laser beams many times
at a single point) or at a linear collider (where the
electron beam would  collide with laser beams at several
points) with reacceleration between them.

The theoretical consideration of the process of multiple Compton
scattering in  papers~\cite{Bes}--\cite{P2} is based on two
different approaches. The first approach  exploits the
fact that the scattered electrons are polarized even if the
initial electron beam is unpolarized. Since the scattered
electrons do not leave the beam, multiple collisions with 
laser photons would appear to lead to a growth of the mean electron
beam polarization. The quantitative consideration of this idea in
papers~\cite{Ros,P1} resulted in
 \begin{itemize}
 \item {\bf Conclusion 1}~\cite{Ros,P1}. {\it A longitudinal
    polarization of electrons (positrons) up to  $100~\%$ can be
     achieved in a  relatively short time.}
 \end{itemize}
In Sect.~3 we explain the origin of the mistake
that  lead to  ``Conclusion 1''. Briefly speaking, in 
simulation of  multiple Compton scattering one should not only
consider the multiple Compton scattering of the same electron
but also take into account the fact that the
polarization of {\it unscattered} electrons   changes  in the
laser wave as well. The correct simulation procedure for  multiple
Compton scattering (Sect.~2)  leads to zero polarization of the
final electron beam.

In the second approach,  only the equilibrium
polarization of the electron beam after  multiple passes through the
laser beam was considered. Let $w_{+-}$ and $w_{-+}$ be the probabilities for
Compton scattering with a given electron spin-flip. It is not
difficult to show (see Sect.~2.2) that the electron beam gets the
maximal equilibrium polarization $\zeta_z^{(f)}
=(w_{-+}-w_{+-})/(w_{-+}+w_{+-})$. The corresponding probabilities
have been calculated in papers~\cite{Bes,P2} with the following
 \begin{itemize}
\item {\bf Conclusion 2}~\cite{Bes,P2}. {\it The longitudinal
    polarization of electrons (positrons) as large as $62.5\,\%$ can
    be achieved.}
 \end{itemize}
In Sect.~4 we show that  Conclusion 2 is due to an error in the
calculation of the Compton spin-flip cross sections. The error is
connected to the incorrect transition between the collider
frame (CF) and the rest frame of the initial electron (RFIE).
The correct result corresponds to $w_{+-}=w_{-+}$, therefore, the
discussed process of laser polarization is impossible.

Below, in Sect.~2, we present a set of formulae for Compton scattering
that takes into account the particle polarization from
Ref.~\cite{KPS}, as well as a short description of the simulation
procedure for the multiple Compton scattering from
Ref.~\cite{KPerltS}, which are useful for quantitative consideration
of this method. Here we show that scattering of laser photons do not
lead to polarization of electron beams, in fact they lead only to
depolarization. In Sect.~3 and 4 we explicitly show the origin of
mistakes in the previous papers on this subject.

\section{Polarization of final electrons }

\subsection{Polarization of final electrons in single Compton
scattering}

We consider the basic Compton scattering
 \be
 e(p) +\gamma (k) \to e(p') +\gamma (k')
 \label{1}
 \ee
in the CF, in which an electron with energy $E\sim 1$ GeV collides
a head-on  with a laser photon of energy $\omega \sim
1$ eV.

Let us introduce some notation related to the Compton scattering
(\ref{1}) in  CF. We choose the quantization axis ($z$-axis)
along the initial electron momentum ${\bf p}$ (i.e., anti-parallel
to the laser photon momentum ${\bf k}$). Let $P_c=+1$ be the mean
helicity of the circularly polarized laser photons, {\boldmath
$\zeta$}$= (0,\,0,\,\zeta_z)$ and {\boldmath $\zeta
  $}$^{\prime}=(0,\,0,\,\zeta'_z)$ be the polarization vectors of
the electron in the initial and final states. It is convenient to
describe the Compton scattering by the invariants
 \be x={2pk\over
  m_e^2}, \;\;\;\; y={2pk'\over pk}= {2p'k \over pk}.
 \ee
In CF, we have
 \be
  x = {4E\omega\over m_e^2}, \,\,\, y= {\omega'\over E}\,.
 \label{5}
 \end{equation}
The maximum energy of scattered photons is
\be
\max\{ \omega'\} = y_m E\,, \;\;\; y_m = \frac{x}{x+1}\,.
 \label{6}
 \end{equation}
For $E \sim 1$ GeV and $\omega \sim 1$ eV, the value of $x \sim 0.015$,
therefore hereafter we assume 
\be x\ll 1, \,\;\;\;\;\;\;\;\; E-E'=\omega' \ll E\,.
 \label{7}
 \ee

In RFIE, the energy of the laser photon $xm_e/2$ is small in
comparison with the electron mass: $xm_e/2 \ll m_e$; and therefore
the transverse momenta are small as well:
 \be
 |{\bf p}'_\perp| =|{\bf k}'_\perp| < x m_e/2 \ll m_e\,.
 \ee
As a result, the electron scattering angle in  CF is very small
 \be
 \theta_e \approx {|{\bf p}'_\perp|\over  p'_z} < \frac{xm_e}{2E}
 \approx {2\omega \over m_e} \sim 4 \times 10^{-6}\,.
 \label{55}
 \ee

The Compton cross section in the collider frame for the above conditions
has been given in~\cite{KPS}:
 \be
 {d \sigma\over dy} =
{\pi r_e^2\over x}\,\left[(1+\zeta_z\zeta_z')\,F_1
+\left(\zeta_z+\zeta'_z\right)\,P_c\,F_2
+\zeta_z\zeta'_z\,F_3\right]\,,
 \label{10}
 \ee
 $$
 F_1={1\over 1-y}+c^2-y\,,\;\;
 F_2={y(2-y)c \over 1-y}\,,\;\;
 F_3=-{y^2s^2 \over 1-y}\,,
 $$
where $r_e$ is the classical electron radius and
 \be
s=2\sqrt{r(1-r)}\,,\;\;\; c= 1-2r\,,\;\; r={y\over (1-y)x}
 \label{4}
 \ee
(in RFIE, the invariant $s=\sin{\vartheta}$ and the invariant
$c=\cos{\vartheta}$, where $\vartheta$ is the photon scattering
angle).\footnote{Strictly speaking,  Eq.~(\ref{10}) given in
\cite{KPS} is valid for the case when $\zeta'_z$ is the projection
of the vector {\boldmath $\zeta$}$^{\prime}$ on the momentum of
the final electron ${\bf p}'$. However, the projections of
{\boldmath $\zeta$}$^{\prime}$ on ${\bf p}'$ and on the $z$-axis
coincide in our approximation due to the very small value of the
electron scattering angle in CF (\ref{55}).}

\subsection{Balance equations}

If we do not take into account the transverse polarization of
electrons, we can describe the electron beam as a mixture of $N_+$
electrons with $\zeta_z=+1$ and $N_-$ electrons with $\zeta_z=-1$.
In this case the mean value of the longitudinal electron
polarization is
 \be
 \zeta_z = {N_+ -N_-\over N_+ +N_-}\,.
 \label{zeta}
 \ee
When the electron beam travels the path $dz$ in the laser beam
with the laser photon density $n_L(z)$, the change of numbers
$N_{\pm}(z)$ is given by the balance equations derived by the
following simple consideration. A reduction of the number of
electrons  with $\zeta_z=+1$ in the beam is determined by the quantity
$N_+ (z)( dw_{++}+dw_{+-})$, where
\be
 dw_{\zeta_z\,\zeta'_z} = 2\sigma(\zeta_z,\,\zeta'_z) n_L(z) dz
 \ee
is the probability that an electron with a certain $\zeta_z$ is
scattered on the path $dz$ with the transition to a certain
$\zeta_z'$ (here we assume also a certain $P_c=+1$). The
coefficient $2$ is due to the fact that the electron and the laser
photon travel towards each other with the speed of light. On the
other hand, the sum $N_+ (z) dw_{++}+N_- (z) dw_{-+}$ represents
the number of scattered electrons with $\zeta'_z =+1$.

As a result, the total change of the number of electrons with
$\zeta_z =+1$ is equal to
 \be
 dN_+(z) = -N_+ (z) (dw_{++}+dw_{+-})+N_+ (z) dw_{++}+N_- (z)
 dw_{-+}
 \label{b1}
 \ee
 $$
 = -N_+ (z) dw_{+-}+ N_- (z)d w_{-+}
  $$
and, similarly,
\be
 dN_-(z) =  -N_- (z)d w_{-+}+ N_+ (z)d w_{+-}\,.
 \label{b2}
 \ee
Since in the considered method the scattered electrons remain in
the beam, the sum $N_+ +N_-=N_e$ does not change:
 \be
 dN_+(z)+dN_-(z)=0\,,
 \label{b3}
 \ee
while the mean longitudinal polarization (\ref{zeta}), generally
speaking, changes as
 \be
 N_e d \zeta_z =dN_+ (z)-dN_- (z) =-2N_+ (z)d w_{+-}+ 2N_- (z)
d w_{-+}\,.
 \label{b4}
 \ee

Balance equations (\ref{b1})--(\ref{b4}) are simplified in two
particular cases. First, if the initial electron beam is
unpolarized, $N_+ (z)=N_- (z)=N_e/2$, it gets (after traveling the
path $dz$) the polarization
 \be
 d\zeta_z = dw_{-+} -dw_{+-}\,.
 \ee
Second, let us consider the equilibrium polarization of the
electron beam (which is achieved after multiple passing of the
electron beam through the laser beam). In this case, $dN_+ =dN_-
=0$ or
 \be
 N_+\,d w_{+-}= N_-\,d w_{-+}\,.
 \ee
From this equation one obtains the equilibrium polarization degree
of the electron beam
 \be
  \zeta_{z}^{(f)}  = { w_{-+}- w_{+-}\over  w_{-+}+ w_{+-}}
 \ee
where $w_{\pm \mp} =\int dw_{\pm \mp}$ is the probability for the
whole path in the laser beam. The quantity $ w_{+-}- w_{-+}$ is
proportional to the difference of the Compton spin-flip cross
sections (with a certain $P_c=+1$ and with the summation over spin
states of the final photon):
 \be
 \Delta \sigma = \sigma (\zeta_z=+1,
\zeta'_z=-1) - \sigma (\zeta_z=-1, \zeta'_z=+1)\,,
  \label{14b}
  \ee
Therefore, in both of these cases the crucial question is
whether $\Delta \sigma$ is equal to zero or  not equal to
zero.

In papers~\cite{Bes,P2} it was claimed that the difference
 \be
 \Delta \sigma\neq 0\,,
 \label{15b}
 \ee
the process of laser polarization of particles is possible and $
\zeta_{z}^{(f)} = 5/8$. In contrast, using formulas (\ref{10}) we
immediately obtain
 \be
 \Delta \sigma= 0
 \label{16b}
 \ee
and, therefore, the polarization is zero.\footnote{The similar
remark about the equal spin-flip probabilities in CF was given
(without details) in Ref.~\cite{Saldin}.} The origin of the incorrect
result~(\ref{15b}) is explained in Sect.~4.

Taking into account that $dw_{+-}=dw_{-+}$, we can rewrite
Eq.~(\ref{b4}) in the form \be {d\zeta_z \over \zeta_z} =
-4\sigma(\zeta_z=+1,\, \zeta'_z=-1) n_L(z)\, dz \ee from which it
follows that the polarization $|\zeta_z|$ is reduced after traveling
the path $dz$.

Note that the result (\ref{16b}) is due to the specific structure of
Eq.~(\ref{10}): the coefficient in front of $\zeta_z P_c$ in this
equation precisely coincides with the coefficient in front of $\zeta'_z
P_c$.

\subsection{A scheme for simulation of  multiple Compton scattering}

In some problems such as conversion of electrons to photons at
photon colliders, laser cooling, etc., it is necessary to
calculate beam parameters after  multiple acts of Compton scattering.
Let an electron beam traverses a region where  laser light is
focused. It is clear that the energies of these electrons as well
as their polarizations vary due to Compton scattering.

However, when the electron passes through the laser beam, the polarization
varies also for those electrons which conserve their energies and
directions of motion (unscattered electrons). This effect is due
to the interference of the incoming electron wave and the electron
wave scattered at zero angle. The change in the electron
polarization depends not only on the Compton cross section but on
the real part of the forward Compton amplitude as well. Such an
effect was considered in Ref.~\cite{KPerltS}.

Both of these effects should be taken into account in simulation of
 multiple Compton scattering. It can be taken into account in the
following way. The electron state is defined by the current values
of its energy $E$, the direction of its momentum (along the $z$ -axis)
and its mean polarization vector $\mbox{\boldmath $\zeta$}$. The
probability to scatter on the path $dz$ is equal to
 \be
  dw\, = 2\, \sigma (E,\, \zeta_z)\, n_L(z)\, dz\,,
   \ee
where $\sigma(E,\, \zeta_z)$ is the total cross section of the
Compton scattering process. Then, as usual, one can simulate
whether the scattering takes place on this path $dz$ or not.

If the scattering does take place, then, using known formulae for the
Compton cross section in  CF (see Ref.~\cite{KPS}), one can
calculate a new value of the electron polarization vector
$\mbox{\boldmath $\zeta$}^{(f)}$ and other parameters.

If the scattering does not occur, one still has to change the
electron polarization vector.\footnote{The necessity of this step can
  also be seen from the following consideration. The value of the
  Compton cross section depends on polarizations of electron and laser
  beams. If the electron beam was initially unpolarized, then, after
  the Compton scattering of one electron, the rest (unscattered) part
  of the beam gets some polarization (see Sect.~2.2).  That is just
  because electrons with different polarizations have different
  scattering probabilities. In other words, the laser beam ``selects''
  preferably electrons with a certain polarization.  In particular,
  equation (\ref{z}) for the longitudinal polarization can be obtain
  from the balance equations discussed above.} The variation of
electron polarization in the laser wave for a general case was
considered in ~\cite{KPerltS}. Following that paper, the change of the
electron polarization vector of the unscattered electron is
\begin{eqnarray}
 d\zeta_x &=& (R \zeta_y +I \zeta_z \zeta_x) \,P_c\,2\pi r^2_e \, n_L
 dz\,,
\label{x}\\
 d\zeta_y &= &(-R \zeta_x +I \zeta_z
\zeta_y)\,P_c\,2\pi r^2_e \, n_L dz\,,
 \label{y} \\
\label{z}
 d\zeta_z &= &- I (1-
\zeta_z^2)\, P_c \,2\pi r^2_e \, n_L dz \,,
\end{eqnarray}
where the functions $I=I(x)$ and $R=R(x)$ are equal to:
\begin{eqnarray}
I &=& {2\over x}\,\int_0^{x/(1+x}\, {y(2-y)c\over 1-y}\, dy =
 \\
 & = & {2\over x} \left[ \left( 1+ {2\over x} \right)\, \ln{(x+1)}
- {5\over 2}+ {1\over x+1} - {1\over 2(x+1)^2} \right] \,,
 \nonumber
\end{eqnarray}
\begin{eqnarray}
R(x)&=& {2\over \pi\, x} \left[
         \left( 1- {2\over x} \right) F( x-1)
       - \left( 1+ {2\over x} \right) F(-x-1)-\right. \nonumber \\
  & & \left. -{2x^3 \ln{x} \over (x^2-1)^2}
+ {x\over x^2-1}-{2\pi^2\over 3x} \right]  , \label{22}
\end{eqnarray}
with
\begin{equation}
F(x) \, = \, \int\limits_0^x \, {\ln{|1+t|}\over t} \, dt
\end{equation}
being the Spence function.

Now we can calculate the total change of the electron beam
polarization on the distance $dz$ in the laser target by the means of
simulation.

If the initial electron is unpolarized $(\zeta_z=0)$ and the laser
photon is circularly polarized $(P_c=+1)$, then from Eq.~(\ref{10})
we have
 \be
  \sigma = {1\over 2}\left[ \sigma_{\rm unpol} +\pi
r_e^2\,I(x)\,
  \zeta'_z \right]\,,
 \label{12b}
 \ee
where $\sigma_{\rm unpol}$ is the Compton cross section for
unpolarized beams. Therefore, the scattered electron becomes
polarized after the first scattering, and its mean degree of
polarization is
 \be
  \zeta^{(f)}_z = {\pi r_e^2\,I(x)\over \sigma_{\rm unpol}}\,.
 \label{13b}
 \ee
The total number of scattered electrons on the path $dz$ in the
laser target is
 \be
 dN_e = 2\,\sigma_{\rm unpol}\, N_e n_L dz\,.
 \ee
This means that the sum of polarizations of the scattered electrons 
along the $z$-axis:
 \be
\zeta^{(f)}_z\, dN_e = 2\pi r_e^2 \,I(x) N_e n_L dz\,.
 \ee

On the other hand, the unscattered electrons become polarized in
accordance with Eq.~(\ref{z}), and their sum of polarizaions along
the $z$-axis
 \be
d\zeta_z\, N_e =- 2\pi r_e^2 \,I(x) N_e n_L dz\,,
 \ee
i.e. it completely compensates the above polarization of the
scattered electrons. So, {\it the electron beam remains unpolarized}.
This coincides with our  result  obtained  in Sect.~2.2 where only
the equilibrium final state was considered.

At the end of this subsection we note the following. The
components of the vector $\mbox{\boldmath $\zeta$}$ define the
parameters of the polarization density matrix of an electron,
among them, $\zeta_z$ is related to the diagonal matrix elements
while $\zeta_{x}$ and $\zeta_{y}$ determine the off-diagonal
matrix elements. If one is interested in the longitudinal electron
polarization only (as in the present paper), it is sufficient in
simulation to use Eq.~(\ref{z}) connected with the ``occupancy''
numbers $N_{\pm}$. In the general case one should use the whole set of
Eqs.~(\ref{x})--(\ref{z}).

\section{Remark on Conclusion 1}

Now we are ready to show the origin of the error in Conclusion 1.
Let us describe the procedure of the naive simulation of the
multiple Compton scattering. We consider the case when in the
CF the polarization vectors of the initial and final electrons
have $z$-components only and the parameter $x$ is small. The
corresponding Compton cross section with an accuracy up to the terms
of the order of $x$ can be easily obtained from (\ref{10}):
 \be
\sigma ={4\over 3} \pi r_e^2\,\left[ (1-x)(1+ \zeta_z\,\zeta'_z) -
{x\over 4}\, P_c (\zeta_z+\zeta'_z)\right]\,.
 \label{11}
 \ee
 If the initial
electron is unpolarized $(\zeta_z=0)$ and the laser photon is
circularly polarized $(P_c=+1)$, then
 \be
\sigma ={4\over 3} \pi r_e^2\,\left[ 1-x - {x\over 4}\, \zeta'_z
\right]\,,
 \label{12}
 \ee
i.e. the cross section is somewhat larger for $\zeta'_z =-1$ than
for $\zeta'_z =+1$. Therefore, the scattered electron becomes
polarized after the first scattering and its mean degree of 
polarization is
\be
  \zeta^{(f)}_z = - {x\over 4}\,.
 \label{13}
 \ee
Repeating this procedure, one can find that after $N$
collisions the electron polarization is
 \be
  \zeta^{(f)}_z = -  {N\over (4/x)+N}\, ,
 \label{14}
 \ee
which can reach $100\, \%$ for $N\gg 4/x$. This fact is the basis
for Conclusion 1.

This ``polarization'' is not connected with the electron
spin-flip, it is due to some differences in the cross sections: the
polarized laser beam selects electrons with a certain (in our
case, negative) polarization. But such a naive simulation of the
multiple Compton scattering is incorrect because it does not take
into account the fact that unscattered electrons become polarized
in the opposite direction. The correct procedure for this simulation is
described in the previous section and leads to zero polarization.

\section{Remark on Conclusion 2}

In Sect.~2.1 we have shown that the equilibrium polarization of
electrons in the discussed method is zero. Below, we show the
origin of the mistake that led to Conclusion 2. We  remind that
our result (\ref{16b}) has been obtained in the CF. To the
contrary, the authors of Conclusion 2 had obtained their result
(\ref{15b}) in the RFIE. Below we demonstrate how to obtain our
result in RFIE and show that the error in Conclusion 2 is
connected with inaccurate transition from  CF to  RFIE.

In our consideration, we use the electron polarization
vectors\footnote{ They determine the electron-spin 4-vectors $a$
and $a'$ as
 \begin{equation}
a=\left({\mbox {\boldmath $\zeta$}{\bf p}\over m_e}\,,\,
\mbox{\boldmath $\zeta$}+ {\bf p}{\mbox{\boldmath $\zeta$} {\bf p}
\over m_e(E+m_e)}\right) \,,\;\;\;
 a^{\prime}=\left({\mbox
{\boldmath$\zeta$}^{\prime} {\bf p}^{\prime}\over m_e}\,,\, \mbox
{\boldmath$\zeta$}^{\prime} +{\bf p}^{\prime}{\mbox
{\boldmath$\zeta$}^{\prime} {\bf p}^{\prime}\over
m_e(E^{\prime}+m_e)}\right)\,.
 \label{13a}
\end{equation}
}
{\boldmath $\zeta$} and {\boldmath $\zeta$}$'$, which in CF have the
forms
 \be
 {\mbox{\boldmath $\zeta$}}=(0,\,0,\,\pm 1)\,,\;\;
 {\mbox{\boldmath $\zeta$}}'=(0,\,0,\,\mp 1)\,.
 \label{14a}
 \ee
It is not difficult to show that in  RFIE the vector {\boldmath
$\zeta$} has the same form, but the vector {\boldmath $\zeta$}$'$
has another form:
 \be
  \mbox{\boldmath $\zeta$}'_\perp =
   \mp \,  {{\bf p}'_\perp \over m_e}=
  \pm \,  {{\bf k}'_\perp \over m_e}  \,,\;\;
  \zeta'_z = \mp \,\left( 1 - {({\bf p}'_\perp)^2 \over 2 m^2_e}\right)
  \approx \mp 1  \,,
 \label{15a}
 \ee
since transition from CF to RFIE corresponds to a boost along
the vector ${\bf p}$,  but not along the vector ${\bf p}'$. To
prove (\ref{15a}), it is sufficient to consider the transverse
component of the electron-spin 4-vector $a'$ in the CF and in the
RFIE. In the CF, one has
 \be
{\bf a}'_\perp= \mp\, {\bf p}'_\perp \, {p'_z\over
m_e(E'+m_e)}\approx \mp\, {{\bf p}'_\perp\over m_e}\,,
 \ee
since in the CF we have $p'_z/(E'+m_e)\approx 1$. In  RFIE the
same polarization vector is
 \be
{\bf a}'_\perp= {\mbox{\boldmath $\zeta$}}'_\perp +{\bf p}'_\perp
\, {{\mbox{\boldmath $\zeta$}}' {\bf p}'\over m_e(E'+m_e)} \approx
{\mbox{\boldmath $\zeta$}}'_\perp\,,
 \ee
since in  RFIE we have $|{\bf p}'|/ m_e \ll 1$.

The needed Compton cross section in  RFIE can be found in the
textbook~\cite{BLP} (see Eqs.~(87,22) and (87,23)):
 \be
{d\sigma\over d\Omega}= {r^2_e\over 4}\,\left( {\omega'\over
\omega}\right)^2\, \left[ F_0 +{\bf f}\mbox{\boldmath
$\zeta$}P_c+{\bf g}\mbox{\boldmath $\zeta$}'P_c+G_{ik}\zeta_i
\zeta'_k\, \right]\,.
 \label{16a}
 \ee
 From this we get the following result:
 \be
{d\Delta\sigma\over d\Omega}= {r^2_e\over 2}\,\left( {\omega'\over
\omega}\right)^2\, \left[\, f_z- g_z+ {\bf g}_\perp {{\bf
k}'_\perp\over m_e} \, \right]\,.
 \label{17a}
 \ee
Before proceeding, we note that the momenta of the initial
and final photons in  RFIE are
 \be
{\bf k}=(0,0, -\omega )\,,\;\; {\bf k}'= ({\bf k}'_\perp, \,
-\omega' \cos{\vartheta})
 \label{18a}
 \ee where $\vartheta$ is the photon scattering angle (the direction
 of the $z$-axis is along the vector $(-{\bf k})$). Using Eq.~(87,23) from
 \cite{BLP} and taking into account the terms of the second order in
 $\omega/m_e$ we obtain \be f_z-g_z = {\omega-\omega'
   \cos{\vartheta}\over m_e}\, {\omega+\omega'\over
   2m+\omega-\omega'}\, \sin^2{\vartheta} \approx \left(\omega\over
   m_e\right)^2\, (1-\cos{\vartheta})\, \sin^2{\vartheta}\,,
 \label{19a}
 \ee
 \be
{\bf g}_\perp {{\bf k}'_\perp\over m_e}=- {\left({\bf k}'_\perp
\right)^2\over m_e^2}\, (1-\cos{\vartheta})\,
 \sin^2{\vartheta} \approx  -\left(\omega\over m_e\right)^2\,
 (1-\cos{\vartheta})\,
 \sin^2{\vartheta}\,.
 \label{20a}
 \ee
As a result, we get
 \be {d\Delta\sigma\over d\Omega}=0\,,
 \label{21a}
 \ee
which is in agreement with the conclusion (\ref{16b}) in  CF.

The wrong conclusion (\ref{15b}) was obtained because the
same form (\ref{14a}) was used for the vector {\boldmath
$\zeta$}$'$ both in the CF and in the RFIE. It is equivalent to
omitting the last term in the square bracket in Eq.~({\ref{17a}).

Thus, the calculations, performed in  CF as well as in RFIE,
give us the same result (\ref{16b}). We, therefore, conclude that the
claim (\ref{15b}) is based on an inaccurate transition from
CF to RFIE.

\section{Summary}

We have shown that the multiple Compton scattering of electrons on
circularly polarized laser photons at usual storage rings or linear
accelerators does not lead to polarization of electron beams.
Statements by some authors about obtainability of high degrees of
polarization are explained by mistakes in their calculation
procedures. We had discussions with E.G.~Bessonov and A.P.~Potylitsyn,
and they agreed with our criticism.

In this paper we have considered the {\it linear} Compton
scattering (the scattering of an electron on a single laser
photon). It is technically possible to realize  conditions which
correspond to the {\it nonlinear} Compton scattering (the
scattering of an electron on several laser photons). The effective
cross section for the nonlinear Compton scattering from
Ref.~\cite{Gal} has the same specific structure as Eq.~(\ref{10})
but with much more complicated functions $F_{1,2,3}$. From this,
one can easily obtain the result (\ref{16b}), which means that the
equilibrium polarization of electrons is zero in the the case of
the nonlinear Compton scattering as well.

One additional remark. There is no polarization of the electron
beam as a whole in the considered scheme, however, it does not
close the possibility to use lasers for polarization of electron
beams in other schemes. For example, it has been shown in
Ref.~\cite{Saldin} that using specially arranged spin-orbit
coupling in damping rings (by adding a solenoid), a polarization of
about 60 \% may be reached. This method is based on the difference
in the Compton cross sections for electrons with different values
of their helicities, on the fact that scattered electrons have
lower energy compared to unscattered electrons, and on dependence
of the spin precession angle on the electron energy. This method
is not simple, and is too slow for preparation of beams for
linear colliders.

\section*{Acknowledgement}

We are very grateful to E.~Bessonov, R.~Brinkmann, V.~Katkov,
A.P.~Potylitsyn,  E.L.~Sal\-din, A.N.~Skrinsky and V.~Strakhovenko
for useful discussions.  This work is supported in part by INTAS
(code 00-00679), RFBR (code 00-02-17592 and 00-15-96691) and by
St. Petersburg grant (code E00-3.3-146).

\end{document}